\documentclass[aps,amsmath,amssymb,twocolumn]{revtex4-1}

\usepackage{graphicx}
\usepackage{dcolumn}
\usepackage{bm}
\usepackage{color}
\usepackage[normalem]{ulem}

\begin{document}

\title{Normal Metal-Superconductor Near-Field Thermal Diodes and Transistors}

\author{E. Moncada-Villa$^{1}$}
\author{J.~C. Cuevas$^{2}$}

\affiliation{$^{1}$Escuela de F\'{\i}sica, Universidad Pedag\'ogica y Tecnol\'ogica de Colombia,
Avenida Central del Norte 39-115, Tunja, Colombia}

\affiliation{$^2$Departamento de F\'{\i}sica Te\'orica de la Materia Condensada
and Condensed Matter Physics Center (IFIMAC), Universidad Aut\'onoma de Madrid,
E-28049 Madrid, Spain}

\date{\today}

\begin{abstract}
In recent years there has been a number of proposals of thermal devices operating in the near-field regime that
make use of phase-transition materials. Here, we present a theoretical study of near-field thermal diodes and transistors 
that combine superconducting materials with normal (non-superconducting) metals. To be precise, we show that a system 
formed by two parallel plates made of Nb and Au can exhibit unprecedented rectification ratios very close to unity 
at temperatures around Nb superconducting critical temperature and for a wide range of gap size values within the
near-field regime. Moreover, we also show that a superconducting Nb layer placed between Au plates can operate as a 
near-field thermal transistor where the amplification factor can be greatly tuned by varying different parameters such 
as the temperature and thickness of the Nb layer or the distance between the Nb layer and the Au plates. Overall, our 
work shows the potential of the use of superconductors for the realization of near-field thermal devices.
\end{abstract}

\maketitle

\section{Introduction}

When two objects at different temperatures are separated by a distance smaller than the thermal wavelength given by 
Wien's displacement law ($\sim$10 $\mu$m at room temperature), they can exchange thermal radiation via evanescent waves
(or photon tunneling). Such a contribution to the near-field radiative heat transfer (NFRHT) can dominate the heat exchange
for small gaps and lead to overcome the blackbody limit set by Stefan-Boltzmann's law for the radiative heat transfer between
two bodies \cite{Song2015a,Cuevas2018,Biehs2020}. This NFRHT enhancement was first predicted by Polder and van Hove in the early 
1970's \cite{Polder1971} making use of the so-called theory of fluctuational electrodynamics \cite{Rytov1953,Rytov1989}. In
recent years this idea has been thoroughly tested and confirmed in a great variety of systems and using different types of materials 
\cite{Kittel2005,Narayanaswamy2008,Hu2008,Rousseau2009,Shen2009,Shen2012,Ottens2011,Kralik2012,Zwol2012a,Zwol2012b,Guha2012,
Worbes2013,Shi2013,St-Gelais2014,Song2015b,Kim2015,Lim2015,St-Gelais2016,Song2016,Bernardi2016,Cui2017,Kloppstech2017,Ghashami2018,
Fiorino2018,DeSutter2019}. 

It is safe to say that at this stage the basic physical mechanisms underlying NFRHT are relatively well-understood. For this
reason, efforts in the thermal radiation community are now shifting towards the proposal and realization of novel functional 
devices based on NFRHT. In this regard, a natural research line that is being pursued is the investigation of the near-field 
thermal analogues of the key building blocks of today's microelectronics: diodes, transistors, switches, memory elements, etc.
Thus far, the diode or rectifier has been the most widely studied thermal device. There have been a lot theoretical proposals
to achieve thermal rectification that make use of systems with dissimilar materials that, in turn, exhibit optical properties
that depend on temperature. The proposed material combinations include SiC structures \cite{Otey2010}, doped Si films 
\cite{Basu2011}, dielectric coating \cite{Iizuka2012}, and Si and a different material \cite{Wang2013}, just to mention a few.
However, the most promising proposals for near-field thermal diodes are based on the use of phase-transition materials 
\cite{Ben-Abdallah2013,Yang2013,Huang2013,Nefzaoui2014,Yang2015,Ghanekar2016,Zheng2017}. An ideal example is that of vanadium 
dioxide (VO$_2$), which undergoes a phase transition from insulator below 340 K to a metal above that temperature. 
This phase transition is accompanied by a drastic change in the infrared optical properties, which has a strong impact in the 
corresponding radiative heat transfer in systems featuring this material as the temperature is varied across the transition
temperature \cite{Zwol2011a,Zwol2011b,Ito2017}. In fact, several experiments have already demonstrated rectification between 
VO$_2$ and SiO$_2$ in the far-field regime \cite{Zwol2012a,Ito2017,Ito2014}. More importantly for our work, the first observation 
of thermal rectification in the near-field regime has been recently reported between a Si microdevice and a macroscopic VO$_2$ 
film \cite{Fiorino2018b}. In that work, a clear rectifying behavior that increased at nanoscale separations was observed 
with a maximum rectification ratio exceeding 50\% at $\sim$140 nm gaps and a temperature difference of 70 K. This high 
rectification ratio was attributed to the broadband enhancement of heat transfer between metallic VO$_2$ and doped Si surfaces, 
as compared to the narrower-band exchange that occurs when VO$_2$ is in its insulating state. From the theoretical point of
view, it has been shown that the rectification ratio can be boosted by nanostructuring VO$_2$ films to form, e.g., one-dimensional
gratings \cite{Ghanekar2016,Ghanekar2018} and the highest reported values for the rectification ratio in the near-field reach 
about 94\% \cite{Ghanekar2016}. 

In 2014, Ben-Abdallah and Biehs extended the idea of thermal diodes based on a phase-transition material to propose the
realization of a near-field thermal transistor \cite{Ben-Abdallah2014}. Their proposed transistor featured a three-body system 
(two diodes in series) in which a layer of a metal-to-insulator transition material (the gate) is placed at subwavelength 
distances from two thermal reservoirs (the source and the drain). In this device, the temperatures of the reservoirs are 
kept fixed, while the temperature of the gate is modulated around its steady-state temperature. Making use of an extension
of the theory of fluctuational electrodynamics to deal with extended many-body systems \cite{Messina2012}, Ben-Abdallah and 
Biehs showed that by changing the gate temperature around its critical value, the heat flux exchanged between the hot body 
(source) and the cold body (drain) can be reversibly switched, amplified, and modulated by a tiny action on the gate. Let us
also say that these ideas have been extended to propose other key elements such as a thermal memory \cite{Kubytskyi2014}, and 
it has also been shown that thermal logic gates can be realized exploiting the near-field radiative interaction in $N$-body 
systems with phase-transition materials~\cite{Ben-Abdallah2016}. These proposals are nicely reviewed in 
Refs.~\cite{Ben-Abdallah2017,Biehs2020}.

\begin{figure}[t]
\includegraphics[width=\columnwidth,clip]{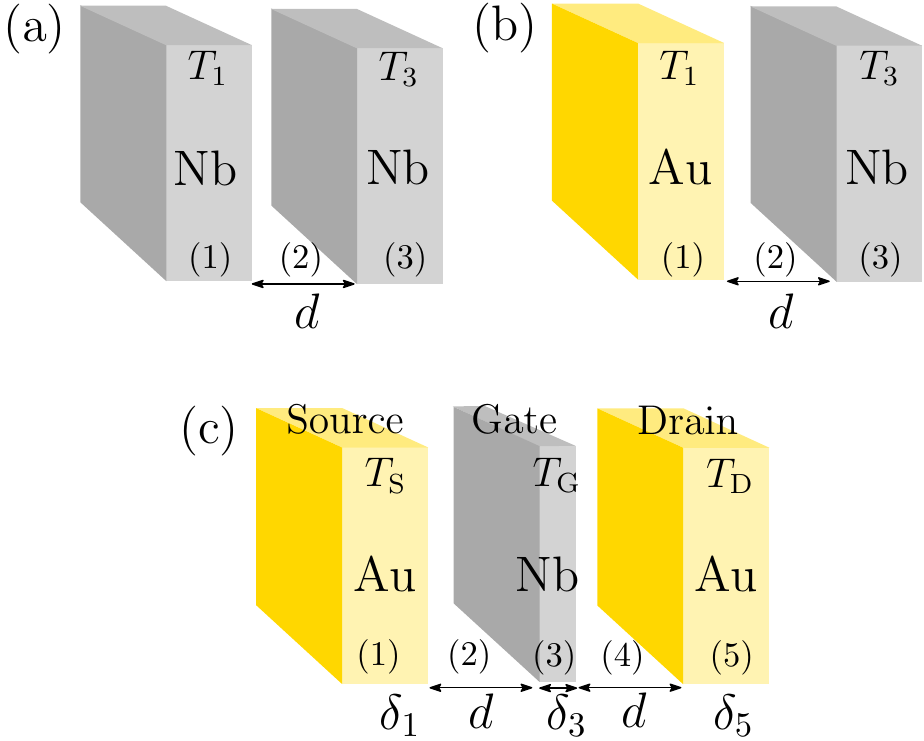}
\caption{Schematic representation of the three systems considered in this paper. The first one [panel (a)] consists of two 
infinite plates of a superconducting material as Nb, each one with its respective temperature $T_1$ and $T_3$, and separated 
by a vacuum gap of size $d$. The second one [panel (b)] is a rectifier composed of a Nb infinite plate exchanging radiative 
heat with an infinite Au plate separated by a distance $d$. Finally, in panel (c), we display a near-field thermal superconducting
transistor. Source and drain, each one with fixed temperatures $T_\mathrm{S}$ and $T_\mathrm{D}$, respectively, are assumed 
to be two infinite Au slabs ($\delta_1 = \delta_5 = \infty$), whereas the gate is made of Nb, which undergoes a 
normal-superconducting phase transition at a critical temperature, $T_\mathrm{C}$, as its temperature, $T_\mathrm{G}$, 
is varied.}
\label{fig-system}
\end{figure}

Since phase-transition materials are ideally suited for near-field thermal management devices, it is natural to
think of superconductors. Although superconductors require to work at low temperatures, their use has the advantage that 
since the thermal wavelength is inversely proportional to temperature, the near-field regime extends to gaps beyond 
the millimeter scale for temperatures around 1 K. Thus, it is easy to reach gaps or separations for which the 
radiative heat transfer is enhanced beyond the blackbody limit. When a normal metal undergoes a superconducting phase
transition (we focus here on conventional low-temperature superconductors), its optical properties change drastically in the
microwave range due to the appearance of a gap in its density of states (of the order of 1 meV depending on the superconductor). 
This gap reduces the emissivity of the metal and in the superconducting state, one expects a substantial reduction of the NFRHT 
when a second material is brought in close proximity. This naive idea has been experimentally confirmed in recent years with 
measurements of the NFRHT between parallel plates made of superconductors like Nb and NbN \cite{Kralik2017,Musilova2019}. In 
particular, it has been reported that there is a contrast in the NFRHT between the normal and the superconducting state of a 
factor 5 in the case of Nb \cite{Kralik2017} and 8 in the case of NbN plates \cite{Musilova2019}. Inspired by these experiments, 
Ordo\~nez-Miranda \emph{et al.}\ \cite{Ordonez2017} proposed the use of a low-temperature superconductor to realize a near-field
thermal diode. In their proposal, the thermal reservoirs were made of Nb and SiO$_2$ and the device operated at temperatures 
around the superconducting critical temperature of Nb (9 K). In particular, they found that at temperatures 1 and 8.7 K for the
two thermal reservoirs, the rectification factor reached a maximum of 71\% for gaps of the order of 60 $\mu$m, which is quite 
high, but still below what is found in proposals involving vanadium dioxide. One the the goals of this work is to show 
theoretically that the use of normal metals, instead of dielectrics like silica, can boost the performance of near-field thermal 
diodes comprising superconducting materials, even beyond any reported value in VO$_2$-based thermal diodes. To be precise, we shall 
consider an Au-Nb rectifier, see Fig.~\ref{fig-system}(b), and show that rectification ratios very close to unity are achieved in 
a very wide range of gap size values in the near-field regime. On the other hand, we shall also show that a Nb plate (the gate 
reservoir) placed in the middle of a vacuum gap between two Au plates, see Fig.~\ref{fig-system}(c), can behave as near-field 
transistor with amplification factors that can be largely tuned by varying different parameters such as the temperature and 
thickness of the gate or the distance between the gate and the source and drain reservoirs.

The rest of the paper is organized as follows. In Sec.~\ref{sec-systems} we introduce the different systems and devices 
that we analyze in this work and explain how we model the optical properties of the materials involved in these systems (Nb and Au).
In Sec.~\ref{sec-Nb-Nb} we briefly discuss the radiative heat transfer between two parallel plates made of Nb to illustrate
the impact of the superconducting phase transition in the heat exchanged via radiation. Then, Sec.~\ref{sec-diode} is
devoted to the analysis of the radiative heat rectification in a thermal diode made of Au and Nb parallel plates. In 
Sec.~\ref{sec-transistor} we study the operation of a three-body system made of a Nb layer between two Au plates as a 
near-field thermal transistor. Finally, we summarize our main conclusions in Sec.~\ref{sec-conclusions}.   

\section{Systems and optical properties} \label{sec-systems}

As explained in the introduction, the main goal of this work is to study theoretically the performance of near-field 
thermal diodes and transistors that make use of superconducting materials. For this purpose, we shall first briefly analyze 
the impact of the superconducting phase transition in the NFRHT in the case of two Nb parallel plates, see Fig.~\ref{fig-system}(a),
in which we shall consider temperatures below and above the superconducting critical temperature of Nb, $T_\mathrm{C} = 9$ K. The 
near-field diode that we shall investigate is schematically depicted in Fig.~\ref{fig-system}(b) and it consists of two infinite 
parallel plates made of Au and Nb separated by a gap of size $d$. Finally, the near-field transistor that we shall
analyze in detail is shown in Fig.~\ref{fig-system}(c). In this case, a Nb layer of thickness $\delta_3$, referred to as gate, 
is placed in the middle of the vacuum gap between two infinite parallel plates made of Au, referred to as source and drain. 
The temperatures of the source, drain, and gate are denoted by $T_\mathrm{S}$, $T_\mathrm{D}$, and $T_\mathrm{G}$, respectively, 
and we shall assume that $T_\mathrm{S} > T_\mathrm{D}$. The distance between the gate and the source and drain is denoted by $d$ 
and we shall consider temperatures around (both above and below) $T_\mathrm{C}$.

The analysis of the radiative heat transfer in all the cases shown in Fig.~\ref{fig-system} will be done within the 
framework of the theory of fluctuational electrodynamics \cite{Rytov1953,Rytov1989}. In this theory, and within the standard 
local approximation, the optical properties of the materials are fully determined by their frequency-dependent dielectric 
functions (we only consider here nonmagnetic materials). In what follow, we shall describe how we model the dielectric functions 
of the two materials involved in our systems under study, Nb and Au. 

The Nb dielectric function is described here following Ref.~\cite{Zimmermann1991} which, in turn, makes use of
Mattis and Bardeen theory for diffusive superconductors of arbitrary purity \cite{Mattis1958}. This dielectric 
function is given by 
\begin{equation} \label{eq-epsc}
\epsilon_{\rm Nb}(\omega)=\varepsilon_{\infty,{\rm Nb}}+\frac{4\pi i}{\omega}\sigma(\omega),
\end{equation}
where $\varepsilon_{\infty,{\rm Nb}}$ is the high frequency limit for the permittivity, $\sigma(\omega)$ is the optical 
conductivity, $\tau= \sigma_{\rm dc}/(\epsilon_0\omega_\mathrm{p}^2)$ is the relaxation time, $\epsilon_0$ is the vacuum 
permittivity, and $\omega_{\rm p, Nb}$ is the plasma frequency. In the superconducting state, the optical conductivity is 
given by \cite{Zimmermann1991}
\begin{equation} \label{eq-cnsc}
\sigma_{\rm sc}(\omega) = \frac{i\sigma_{\rm dc}}{2\omega\tau} \left( J(\omega) + 
\int_{\Delta}^\infty I_2 \, d\varepsilon\right),
\end{equation}
with
\begin{eqnarray}
\label{eq-j}
J(\omega) & = & \left\{ \begin{array}{ll}
\int^{\hbar\omega+\Delta}_\Delta I_1 \, d\varepsilon, & \hbar\omega \leq 2\Delta \vspace{0.2cm} \\
\int_{\Delta}^{\hbar\omega-\Delta} I_3 \, d\varepsilon + 
\int_{\hbar\omega-\Delta}^{\hbar\omega+\Delta} I_1 \, d\varepsilon, & \hbar\omega \geq 2\Delta
\end{array} \right. , 
\end{eqnarray}
and
\begin{widetext} 
\begin{eqnarray} \label{eqs-I}
I_1 & = &  \left[ \left( 1-\frac{\Delta^2+\varepsilon(\varepsilon-\hbar\omega)}{p_4p_2} \right)\frac{1}{p_4+p_2+i\hbar/\tau} -
\left( 1+\frac{\Delta^2+\varepsilon(\varepsilon-\hbar\omega)}{p_4p_2} \right)\frac{1}{p_4-p_2+i\hbar/\tau} \right]
\tanh\left(\frac{\varepsilon}{2kT}\right), \nonumber \\
I_2 & = &\left[ \left( 1+\frac{\Delta^2+\varepsilon(\varepsilon+\hbar\omega)}{p_1p_2} \right)\frac{1}{p_1-p_2+i\hbar/\tau} - 
\left( 1-\frac{\Delta^2+\varepsilon(\varepsilon+\hbar\omega)}{p_1p_2}  \right)\frac{1}{-p_1-p_2+i\hbar/\tau} \right] 
\tanh\left(\frac{\varepsilon+\hbar\omega}{2kT}\right) + \nonumber \\
& & \left[ \left( 1-\frac{\Delta^2+\varepsilon(\varepsilon+\hbar\omega)}{p_1p_2}  \right)\frac{1}{p_1+p_2+i\hbar/\tau} - 
\left( 1+\frac{\Delta^2+\varepsilon(\varepsilon+\hbar\omega)}{p_1p_2}  \right)\frac{1}{p_1-p_2+i\hbar/\tau} \right] 
\tanh\left(\frac{\varepsilon}{2kT}\right) \nonumber \\
I_3 & = &\left[ \left( 1-\frac{\Delta^2+\varepsilon(\varepsilon-\hbar\omega)}{p_3p_2}  \right)\frac{1}{p_3+p_2+i\hbar/\tau} - 
\left( 1+\frac{\Delta^2+\varepsilon(\varepsilon-\hbar\omega)}{p_3p_2}  \right)\frac{1}{p_3-p_2+i\hbar/\tau} \right] 
\tanh\left(\frac{\varepsilon}{2kT}\right),
\end{eqnarray}
\end{widetext}
where
\begin{eqnarray}
p_1 & = & \sqrt{(\varepsilon+\hbar\omega)^2-\Delta^2}, \nonumber \\
p_2 & = & \sqrt{\varepsilon^2-\Delta^2}, \nonumber \\
p_3 & = & \sqrt{(\varepsilon-\hbar\omega)^2-\Delta^2}, \nonumber \\
p_4 & = & i\sqrt{\Delta^2-(\varepsilon-\hbar\omega)^2}.
\end{eqnarray}
In these expressions, $\varepsilon$ is the energy of the carriers, $\omega$ is the frequency of electromagnetic waves, 
$T$ is the temperature, and $\Delta$ is superconducting gap, whose temperature dependence is approximately described by
\cite{Carless1983}  
\begin{equation}\label{eq-gap}
\Delta(T) = \Delta_0 \left(1-\frac{T}{T_\mathrm{C}}\right)^{1/2}
\left( 0.9663 + 0.7733\frac{T}{T_\mathrm{C}} \right),
\end{equation}
where $T_\mathrm{C}$ is the critical temperature. In the normal state ($T > T_\mathrm{C}$), the corresponding optical
conductivity is given by \cite{Zimmermann1991}
\begin{equation}\label{eq-cnnm}
\sigma_{\rm n}(\omega) = \frac{\sigma_{\rm dc}}{1-i\omega\tau} .
\end{equation}

To describe the Au layers, we use the following Drude-like relative permittivity \cite{Chapuis2008}
\begin{equation}\label{eq-epau}
\epsilon_{\rm{Au}}=\epsilon_{\infty,\rm{Au}}- \frac{\omega^2_{\rm p, Au}}
{\omega (\omega + i \gamma_{\rm{Au}})},
\end{equation}                           
where $\epsilon_{\infty,\rm{Au}}$, $\omega_{\rm p, Au}$ and $\gamma_{\rm{Au}}$ are, respectively, the high frequency limit of
dielectric function, plasma frequency, and damping of the free carrier.

\begin{figure}[h!]
\includegraphics[width=0.9\columnwidth,clip]{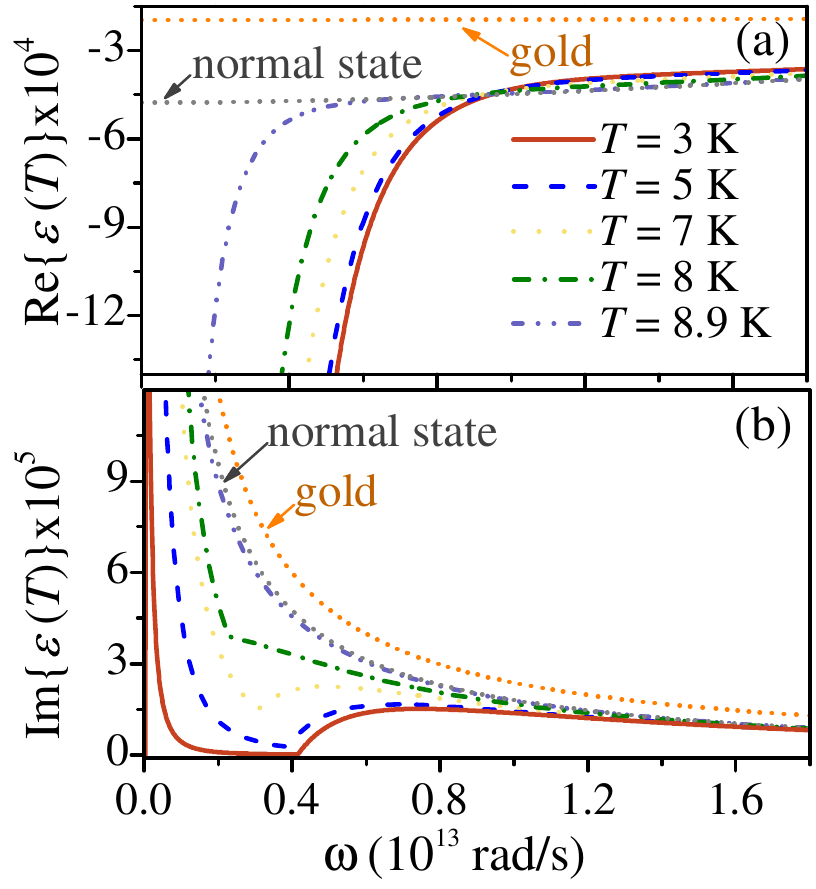}
\caption{Real, panel (a), and imaginary part, panel (b), of the dielectric functions of Nb [see Eq.~\eqref{eq-epsc}] and 
Au [see Eq.~\eqref{eq-epau}] employed in this work. The dielectric function of Nb is shown for different temperatures in 
the superconducting phase, as well as in the normal state.}
\label{fig-perm}
\end{figure}

All calculations in this work were performed with the following parameters for Nb: $\varepsilon_{\infty,{\rm Nb}}=4$
\cite{Zimmermann1991}, $\Delta_0 = 1.764k_\mathrm{B}T_\mathrm{C}$, $T_\mathrm{C} = 9$ K, $\sigma_{\rm dc}=1.7\times 10^7$ 
S/m, and $\omega_\mathrm{p} = 8.8 \times10^{15}$ rad/s \cite{Kralik2017}. For the Au layers, $\epsilon_{\infty,\rm{Au}}=4$, 
$\omega_{\rm p, Au}=1.71 \times 10^{16}$ rad/s, and $\gamma_{\rm{Au}} = 1.22 \times 10^{14}$ rad/s. These parameters are 
consistent with the experimental ones reported in Ref.~\cite{Coste2017} for cryogenic temperatures. In Fig.~\ref{fig-perm}
we display the frequency dependence of the real and imaginary part of the dielectric functions of Nb and Au computed with 
those parameter values. In particular, we show the Nb dielectric function for different temperatures inside the 
superconducting phase, as well as for temperatures above $T_\mathrm{C}$, i.e., in the normal state.

\section{Nb parallel plates} \label{sec-Nb-Nb}

Before discussing the functional devices, diode and transistor, it is convenient to analyze the impact of the superconducting
phase transition in the NFRHT. For this purpose, we revisit here the case of two Nb parallel plates, see Fig.~\ref{fig-system}(a),
which has been analyzed, both theoretically and experimentally, in Ref.~\cite{Kralik2017}. Within the theory of fluctuational
electrodynamics, the net power per unit area (heat flux) exchanged via radiation by two infinite parallel plates, see 
Fig.~\ref{fig-system}(a), is given by \cite{Polder1971}
\begin{equation}
\label{eq-net-Q}
Q = \int^{\infty}_{0} \frac{d \omega}{2\pi} \left[ \Theta_1(\omega) - \Theta_3(\omega) \right] 
\int^{\infty}_{0} \frac{dk}{2\pi} k \left[ \tau_s^{13}+\tau_p^{13}\right] ,
\end{equation}
where $\Theta_i(\omega) = \hbar \omega/ [\exp(\hbar \omega / k_{\rm B}T_i) -1]$, $T_i$ is the absolute temperature of the 
layer $i$, $\omega$ is the radiation frequency, $k$ is the magnitude of the wave vector parallel to the surface planes, and 
$\tau_\beta^{13}(\omega,k,d)$ is the total transmission probability of the electromagnetic propagating ($k < \omega/c$) and 
evanescent ($k> \omega/c$) waves, given by
\begin{eqnarray}
\label{eq-trans-man}
\tau_\beta^{13}(\omega,k,d) = \left\{ \begin{array}{ll}
\frac{ \left( 1 - \left|r_\beta^{21}\right|^2  \right)\left( 1 - \left|r_\beta^{23}\right|^2 \right) } 
{\left|1-r_{\beta}^{21}r_{\beta}^{23}e^{2iq_2d}\right|^2}, & k < \omega/c \vspace{0.2cm}\\
\frac{ 4{\rm Im}\left({r_{\beta}^{21}}\right){\rm Im}\left(r_{\beta}^{23}\right)e^{-2{\rm Im}(q_2)d}} 
{\left|1-r_\beta^{21}r_\beta^{23}e^{2iq_2d}\right|^2}, & k > \omega/c
\end{array} \right. .
\end{eqnarray}
Here, $q_i = \sqrt{\epsilon_i\omega^2/c^2 - k^2}$ is the the wave vector component perpendicular to the plate surfaces in 
the vacuum gap and $c$ is the velocity of light in vacuum. The reflection amplitudes $r_\beta^{ij}$ are given by the Fresnel
coefficients
\begin{equation}
r_s^{ij} = \frac{q_i-q_j}{q_i+q_j} \;\; \mbox{and} \;\;
r_p^{ij} = \frac{\epsilon_j q_i-\epsilon_i q_j}{\epsilon_j q_i+\epsilon_i q_j}.
\end{equation}

The corresponding linear thermal conductance per unit of area, usually referred to as heat transfer coefficient, is given
by  
\begin{equation}
\label{eq-cond}
h = \int^{\infty}_{0} \frac{d \omega}{2\pi} \frac{\partial}{\partial T} 
\left[ \frac{\hbar\omega}{e^{\hbar\omega/k_\mathrm{B}T}-1} \right] 
\int^{\infty}_{0} \frac{dk}{2\pi} k \left[ \tau_s^{13}+\tau_p^{13}\right] .
\end{equation}
In the case of two black bodies, which is achieved when $\tau^{ij}_\beta = 1$ for all frequencies for propagating waves, 
this result reduces to the Stefan-Boltzmann law
\begin{equation}\label{eq-bb}
h_{\rm BB} = 4\sigma T^3,
\end{equation}
where $\sigma$ is the Stefan-Boltzmann constant (not to be confused with a conductivity). 

\begin{figure}[h!]
\includegraphics[width=0.9\columnwidth,clip]{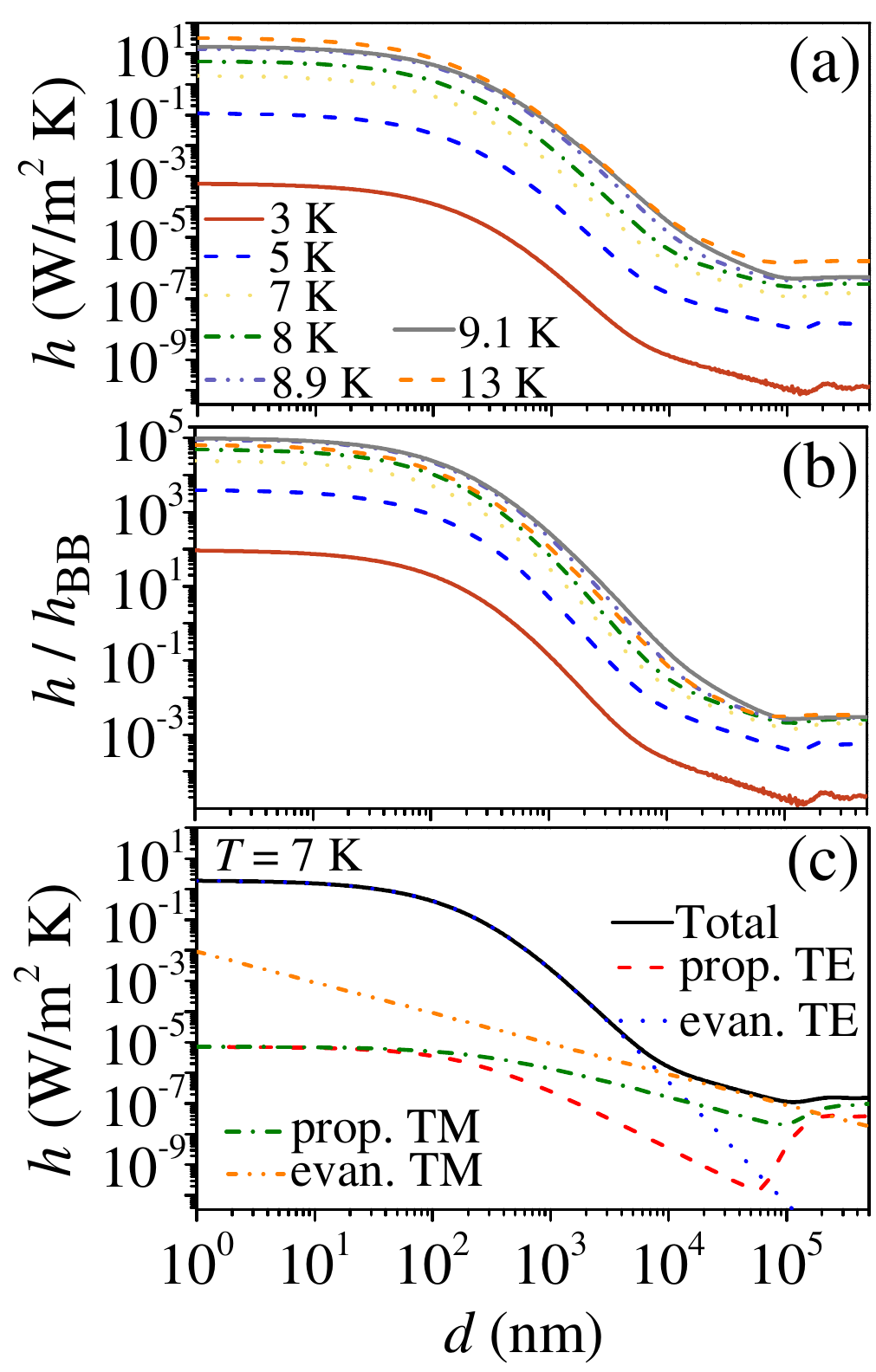}
\caption{(a) Heat transfer coefficient, $h$, for two Nb parallel plates, see Fig. \ref{fig-system}(a), as a function of 
the vacuum gap size $d$. The different curves correspond to different values of the absolute temperature $T$. (b) The
same as in panel (a), but now the heat transfer coefficient is normalized to the corresponding blackbody limit, $h_\mathrm{BB}$ 
[see Eq.~\eqref{eq-bb}]. (c) Different contributions to the heat transfer coefficient of two Nb plates from both 
propagating and evanescent waves with TE and TM polarization and for a temperature of 7 K.}
\label{fig-cond}
\end{figure}

In Fig.~\ref{fig-cond}(a) we show the results for the gap dependence of the heat transfer coefficient of two
Nb parallel plates for different temperatures across the superconducting phase transition ($T_\mathrm{C} = 9$ K).
One can see that the heat transfer coefficient is greatly enhanced in the near-field regime ($d < 10^5$ nm) and it
saturates for small gaps ($d < 100$ nm). Notice also that there is a pronounced temperature dependence, especially
for temperatures below $T_\mathrm{C}$. This strong dependence is due to both the impact of the superconducting phase
transition and the fact that the temperature itself is being changed by an amount comparable to its absolute value.
To disentangle those two dependencies, it is convenient to normalize the heat transfer coefficient by the corresponding
result for two black bodies, as we do in Fig.~\ref{fig-cond}(b). With this normalization, we see that the blackbody 
limit is greatly overcome in the near-field regime (by almost 5 orders of magnitude at $T \sim 10$ K and gaps $d < 100$ nm).
More importantly for this work, there is a very strong reduction of the NFRHT upon decreasing the temperature below 
$T_\mathrm{C}$. For instance, for small gaps the NFRHT for 3 K is about 10 smaller than for $9.1$ K. In simple terms,
this dramatic effect can be explained by the presence of a gap in the spectrum of a superconductor that naturally leads
to a strong reduction of the emissivity of the material at low frequencies, see Fig.~\ref{fig-perm}(b). On the
other hand, and in order to give an insight into the NFRHT in the superconducting phase, we show in Fig.~\ref{fig-cond}(c)
the gap dependence of the different contributions to the heat transfer coefficient for the Nb parallel plates for a 
temperature $T=7$ K, including both propagating and evanescent waves for both TE (or $s$) and TM (or $p$) waves. As expected
for metals, the NFRHT is largely dominated by evanescent TE modes, which can be attributed to total internal reflection 
modes as explained in detail in Ref.~\cite{Chapuis2008}. 

\section{Near-field thermal rectifier} \label{sec-diode}

Now we turn to the analysis of the thermal rectifier shown in Fig.~\ref{fig-system}(b) and composed by two parallel 
plates made of Nb and Au. As explained in the introduction, there have been many proposals for near-field thermal
rectifiers employing a variety of materials that undergo a phase transition as a function of temperature. In particular,
Ordo\~nez-Miranda \emph{et al.}\ \cite{Ordonez2017} proposed a diode with terminals made of parallel plates of Nb and SiO$_2$ 
and operating at temperatures between 1 and 8.7 K, for which Nb is superconducting. These authors reported that the rectification 
factor (see definition below) could reach 71\% for gaps on the order of 60 $\mu$m. In what follows, we show that the performance
of a superconducting near-field thermal rectifier can be boosted by using a metal as a second thermal reservoir instead of a 
polar dielectric like silica. 

The heat flux in an asymmetric system like that of Fig.~\ref{fig-system}(b) can be calculated with the formulas described
in the previous section. For the rectifying behavior, we calculated the net heat flux for both forward (FB) and reverse (RB) 
bias configuration. The forward (reverse) bias heat flux, $Q_{\rm FB}$ ($Q_{\rm RB}$), was calculated by setting $T_1= 1$ K 
and $T_3 = T_1 + \Delta T$ with $\Delta T > 0$ ($T_1 = T_3 + \Delta T$  and $T_3 = 1$ K ), \emph{i.e.}, in forward (reverse) 
bias the temperature gradient is from media (1) to (3) [(3) to (1)]. The rectification factor is defined as
\begin{equation}\label{eq-rect}
\eta = \left| \frac{\left|Q_{\rm FB}\right|-\left|Q_{\rm RB}\right|}
{{\rm max}(\left|Q_{\rm FB} \right|,\left|Q_{\rm RB}\right|)} \right|.
\end{equation}
Notice that with this definition, $\eta$ is bounded between 0 and 1.

\begin{figure}[t]
\includegraphics[width=\columnwidth,clip]{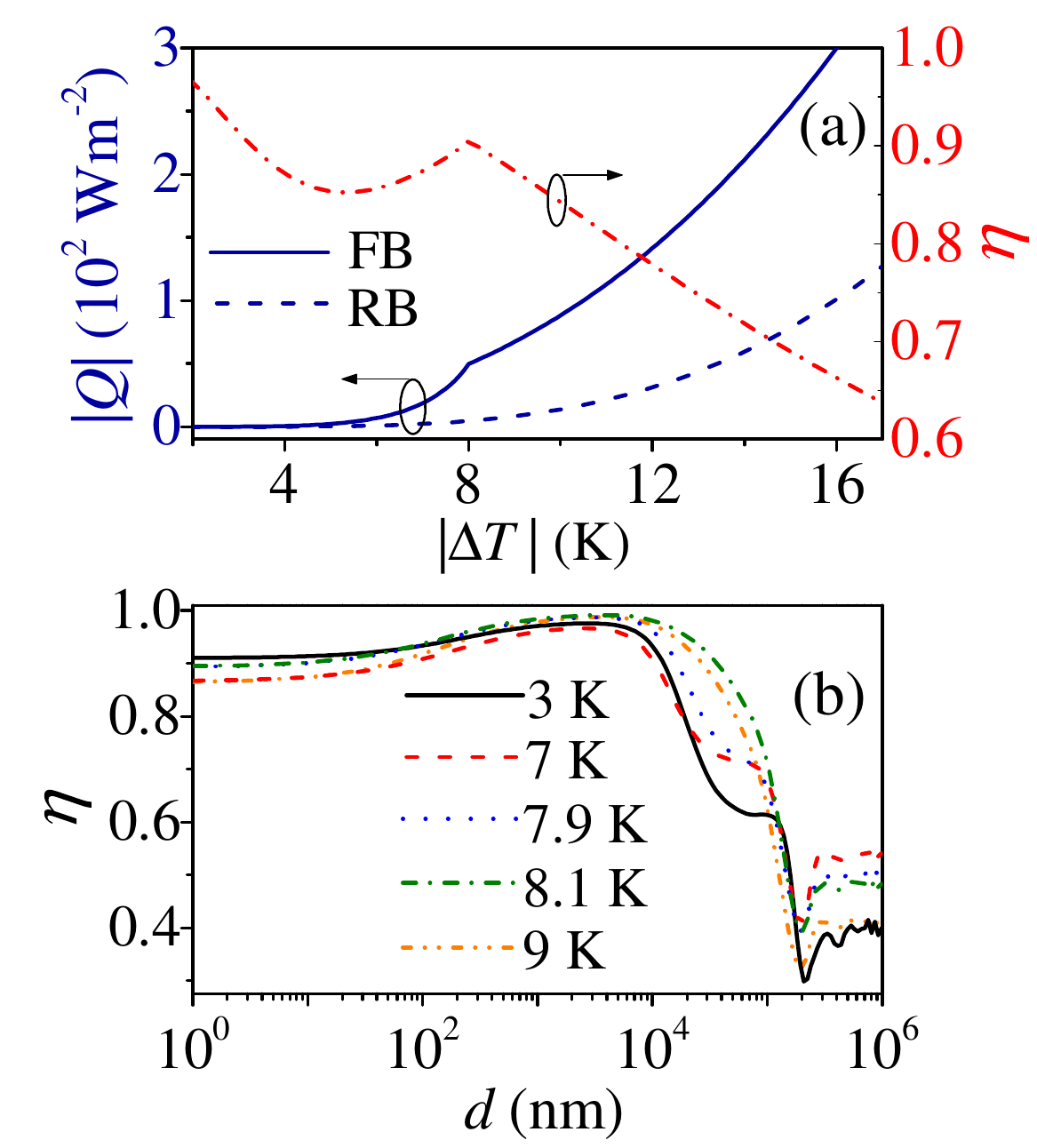}
\caption{(a) Forward (FB) and reverse (RB) bias radiative heat flux in the rectifier of Fig.~\ref{fig-system}(b) as 
a function of the temperature difference $|\Delta T|= |T_1-T_3|$ and for a gap size of 10 nm. The corresponding 
rectification factor $\eta$, defined in Eq. \eqref{eq-rect}, is also included. (b) Rectification factor as a function 
of the vacuum gap size, $d$ and for different values of the temperature difference $|\Delta T|= |T_1-T_3|$. All results 
were obtained for ${\rm min}(T_1,T_3) = 1$ K.}
\label{fig-rectf}
\end{figure}

In Fig.~\ref{fig-rectf} we summarize our main results for the thermal diode of Fig.~\ref{fig-system}(b). In the upper
panel we show the forward and reverse bias heat fluxes for $d=10$ nm as a function of the temperature difference $|\Delta T|$, 
while the temperature of the coldest plate is fixed at 1 K. The corresponding rectification factor is also shown. Notice
that very high values above 0.9 can be achieved for small values of the temperature difference. In the lower panel we
show the results for the gap dependence of the rectification factor for various temperature differences. As one can see,
very high values are reached in the near-field regime in a huge range of gap size values (of about four decades). In particular,
values as high as 98.7\% are obtained for gaps on the order of 10 $\mu$m, see Fig.~\ref{fig-rectf}(b), which to our 
knowledge are the highest ever reported and, in particular, are much higher than those predicted in previous proposals
of superconducting thermal rectifiers \cite{Ordonez2017}.

\begin{figure}[t]
\includegraphics[width=\columnwidth,clip]{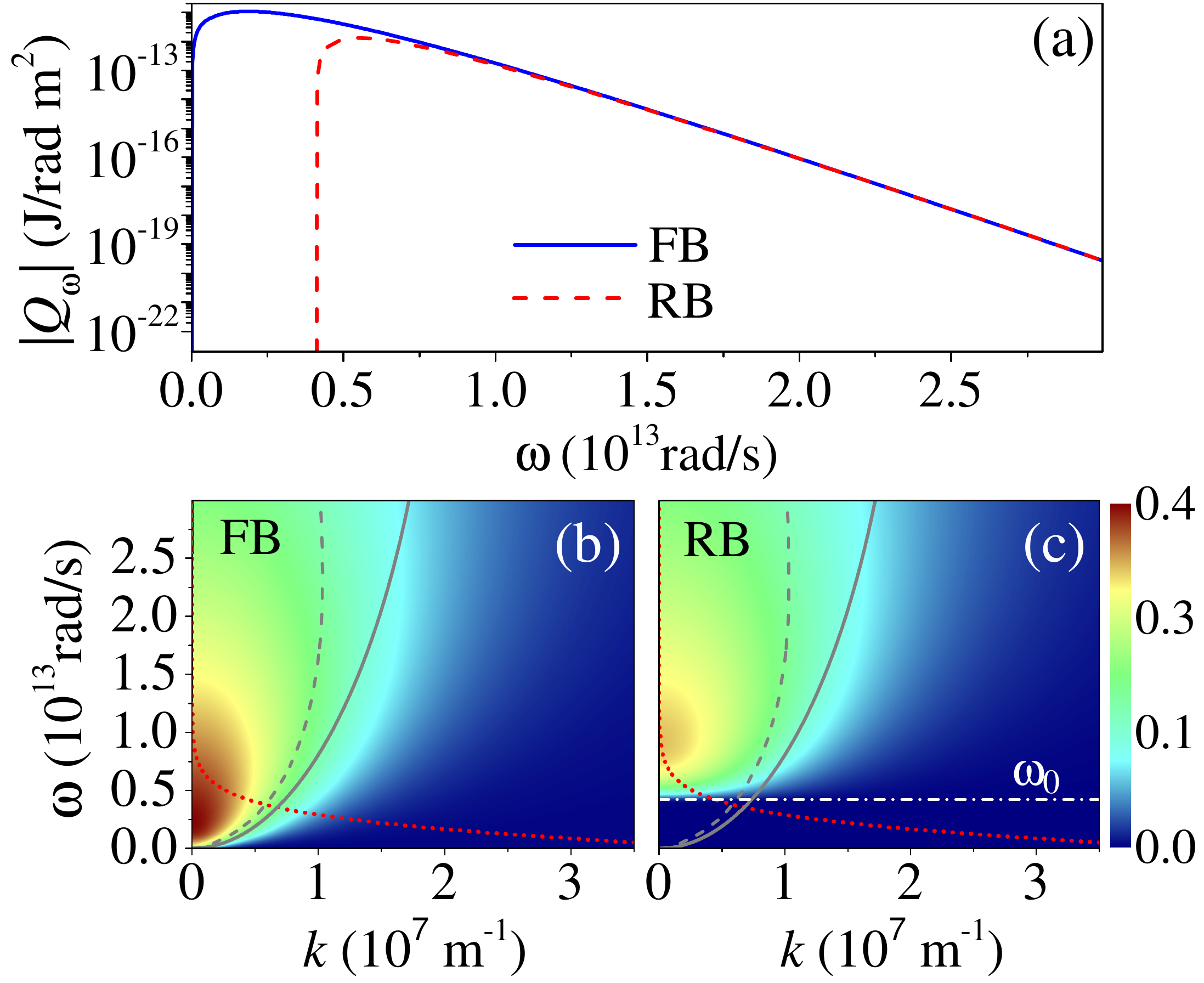}
\caption{(a) Spectral heat flux for the forward (solid line) and reverse (dashed line) bias configurations as a function 
of the frequency and parallel wave vector for a gap size of 10 nm in the system of Fig.~\ref{fig-system}(b). The temperature
difference is set to $|\Delta T| = |T_1-T_3|=7.9$ K, with a temperature of 1 K for the cold reservoir. (b,c) The transmission 
probability of the evanescent TE modes corresponding to each configuration. The dashed and solid lines in both panels represent, 
respectively, the Nb and Au light lines $\omega = ck{\rm Re}(\epsilon_{\rm Nb}^{1/2}) $ and $\omega = ck
{\rm Re}( \epsilon_{\rm Au}^{1/2} )$, whereas the dotted line corresponds to the thermal factor $|\Theta_1-\Theta_3|$ shown
here in arbitrary units. The dash-dotted line in panel (c) indicates the photon frequency required to break a Cooper pair in 
the superconducting state, $\omega_0 = 2\Delta_0/\hbar\approx4.2\times 10^{12}$ rad/s.}
\label{fig-spec}
\end{figure}

To gain some physical insight into the origin of the huge rectification ratios in our system, we show in 
Fig.~\ref{fig-spec}(a) the spectral heat flux (or power per unit of area and frequency) as a function of frequency
for the forward and reverse bias for a case in which the temperatures of the cold and hot reservoirs are 1 and 8.9 K, 
respectively. In this case, the gap size is 10 nm. The spectral heat flux has the characteristic form in metallic systems 
in which the evanescent TE electromagnetic modes completely dominate the NFRHT \cite{Chapuis2008,Kim2015}, which is
also the case in our asymmetric configuration. Notice, in particular, that for the reverse bias there is an abrupt 
frequency cut-off below which the spectral function drastically drops. For the reverse bias, the Nb plate is at 1 K, i.e., 
deep into the superconducting phase, and that cut-off simply corresponds to the frequency $\omega_0 = 2\Delta_0/\hbar
\approx 4.2\times 10^{12}$ rad/s that is required to break a Cooper pair. Below this frequency, the emissivity of 
the Nb plate is drastically reduced due to the presence of a gap in the electronic spectrum, which explains the 
strong reduction of the radiative heat transfer, as compared to the forward bias configuration. This interpretation is further
illustrated in Fig.~\ref{fig-spec}(b,c) where we show the corresponding transmission probability for the evanescent TE modes 
as a function of the frequency and the parallel component of the wave vector. Notice that for the reverse bias, panel (c),
the transmission is very small below $\omega_0$. This fact, together with the frequency dependence of the thermal
factor $|\Theta_1(\omega) - \Theta_3(\omega)|$, see dotted lines in panels (b) and (c), determining the electromagnetic
modes available for heat transfer, see Eq.~(\ref{eq-net-Q}), explains the huge difference between in the net power 
between the reverse and the forward bias configurations.

\section{Near-field thermal transistor} \label{sec-transistor}

Let us now analyze the superconducting thermal transistor depicted in Fig.~\ref{fig-system}(c) that features a Nb
layer of thickness $\delta_3$ as a gate electrode that is placed at a distance $d$ of two Au plates that act as source 
and drain. To compute the different heat exchanges in this three-body system, we have employed the many-body theory put forward 
in Ref.~\cite{Messina2012}. According to this theory, the heat flux received by the drain, $Q_{\rm D}$, and the heat flux 
lost by the source, $Q_{\rm S}$, in the near-field thermal transistor of Fig.~\ref{fig-system}(c) are given by
\begin{eqnarray}
Q _{\rm D} & = & \int^{\infty}_{0} \frac{d \omega}{2\pi} \sum_{\beta=s,p} 
\int^{\infty}_{0} \frac{dk}{2\pi} k \left[\Theta_{13} \tau_\beta^{13}+\Theta_{35} \tau_\beta^{35}\right],  \\ 
Q _{\rm S} & = & \int^{\infty}_{0} \frac{d \omega}{2\pi} \sum_{\beta=s,p} \label{eq-Q_D}
\int^{\infty}_{0} \frac{dk}{2\pi} k \left[\Theta_{53} \tau_\beta^{53}+\Theta_{31} \tau_\beta^{31}\right], \label{eq-Q_S}
\end{eqnarray}
where $\Theta_{ij}=\Theta_i(\omega)-\Theta_j(\omega)$, and $\tau_\beta^{ij}$ is the transmission probability of 
electromagnetic waves from region $i$ to $j$ in Fig.~\ref{fig-system}(c). For a symmetric configuration (source and 
drain equidistant to gate), as considered in this work, these transmissions probabilities are given by 
\begin{eqnarray}
\tau_\beta ^{13} & = & \frac{ 4 \left|\tau_\beta^{3}\right|^2 {\rm Im}\left({\rho_{\beta}^{1}}\right)
{\rm Im}\left(\rho_{\beta}^{5}\right)e^{-4{\rm Im}(q_2)d}} 
{ \left|1-\rho_\beta^{13}\rho_\beta^{5}e^{-2{\rm Im}(q_2)d}\right|^2   
  \left|1-\rho_\beta^{1}\rho_\beta^{3}e^{-2{\rm Im}(q_2)d}\right|^2  } , \nonumber \\
\tau_\beta ^{35}&=& \frac{ 4 {\rm Im}\left({\rho_{\beta}^{13}}\right)
{\rm Im}\left(\rho_{\beta}^{5}\right)e^{-2{\rm Im}(q_2)d}} 
{ \left|1-\rho_\beta^{13}\rho_\beta^{5}e^{-2{\rm Im}(q_2)d}\right|^2 } ,
\end{eqnarray}
where
\begin{eqnarray}
\rho_\beta^{13} & = & \rho_\beta^3 + \frac{ \left(\tau_\beta^{3}\right)^2 \rho_{\beta}^{1} e^{-2{\rm Im}(q_2)d}} 
{  \left|1-\rho_\beta^{1}\rho_\beta^{3}e^{-2{\rm Im}(q_2)d}\right|^2   } , \nonumber \\
\rho_\beta^i & = & - r_\beta^{i}\frac{1-e^{2iq_i\delta_i}}{1-(r_\beta^{i})^2e^{2iq_i\delta_i}}, \nonumber \\
\tau_\beta^i & = & \frac{t_\beta^i\overline t_\beta^ie^{iq_i\delta_i}}{1-(r_\beta^i)^2e^{2iq_i\delta_i}},\nonumber \\
r_s^i & = & \frac{q_2-q_i}{q_2+q_i}, \;\; r_p^i = \frac{\epsilon_iq_2-q_i}{\epsilon_i q_2+q_i}, \nonumber \\
t_s^i & = & \frac{2q_2}{q_2+q_i}, \;\; t_p^i = \frac{2\sqrt{\epsilon_i}q_2}{\epsilon_i q_2+q_i}, \nonumber \\
\overline t_s^i & = & \frac{2q_i}{q_2+q_i}, \;\; \overline t_p^i = \frac{2\sqrt{\epsilon_i}q_i}{\epsilon_i q_2+q_i}. \nonumber
\end{eqnarray}
Additionally, the transmission probabilities $\tau_\beta ^{53}$ ($\tau_\beta ^{31}$) can be obtained after substitution 
$1\rightarrow 5$ and $5\rightarrow 1$ in $\tau_\beta ^{13}$ ($\tau_\beta ^{35}$). The corresponding net heat flux received 
or emitted by the gate is defined as \cite{Ben-Abdallah2014} 
\begin{equation}
Q_{\rm G}=\left|Q_{\rm S}\right|-\left|Q_{\rm D}\right|,
\end{equation}
and the amplification factor as
\begin{equation}
\label{eq-alpha}
\alpha=\left|\frac{\partial |Q_{\rm D}|}{\partial Q_{\rm G}} \right| = \frac{1}{|1- Q^{\prime}_\mathrm{S}/
Q^{\prime}_\mathrm{D}|} ,
\end{equation}
where $Q^{\prime}_\mathrm{S/D}= \partial |Q_{\rm S/D}|/ \partial T_{\rm G}$. Amplification requires a negative differential
conductance and it is characterized by $\alpha > 1$ \cite{Ben-Abdallah2014}. In what follows, we shall choose $T_\mathrm{S} = 
20$ K, $T_\mathrm{D} = 1$ K, and $T_\mathrm{S} > T_\mathrm{G} > T_\mathrm{D}$.

\begin{figure}[t]
\includegraphics[width=\columnwidth,clip]{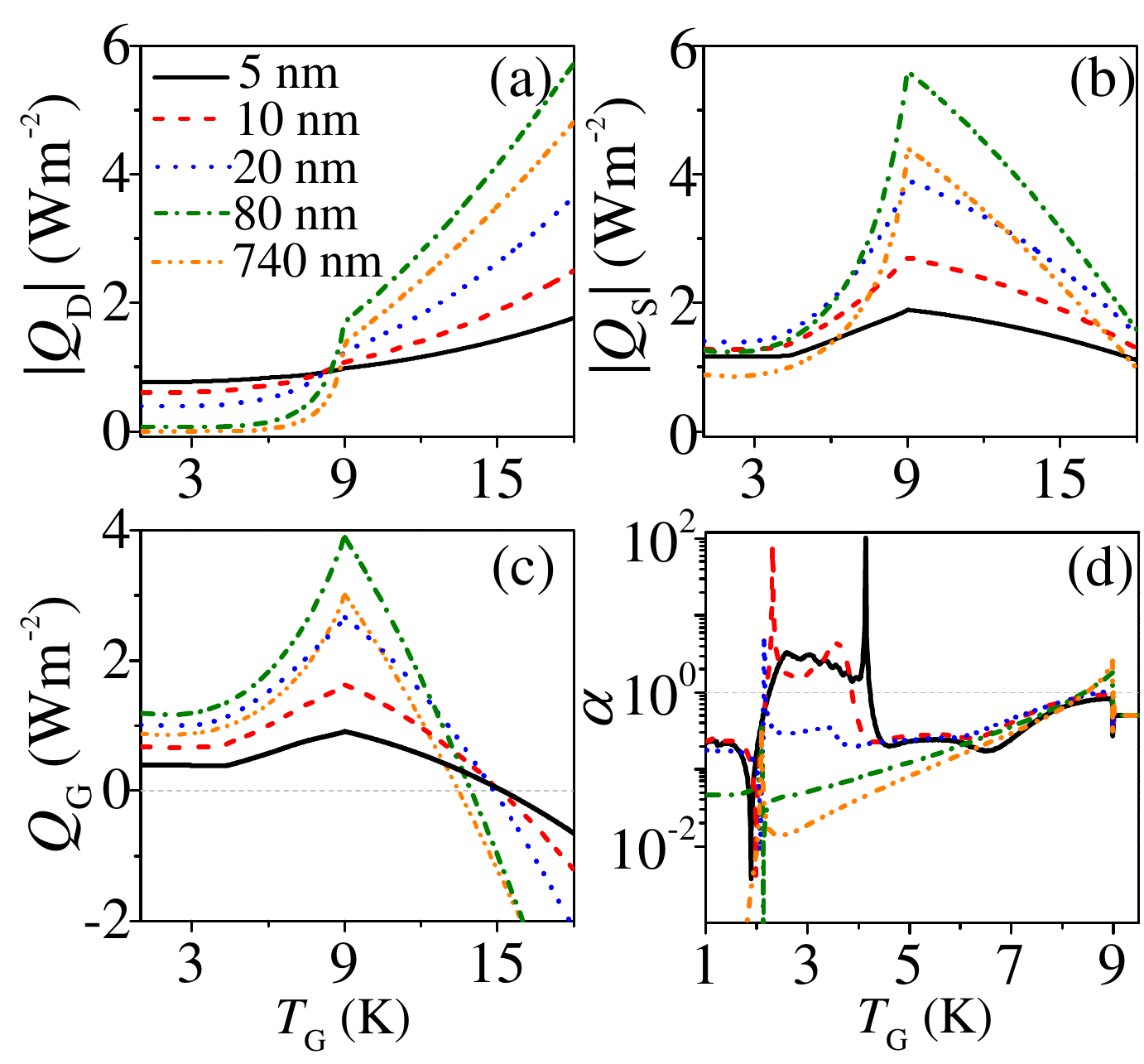}
\caption{(a) Absolute value of the of the radiative heat flux received by the drain in the thermal transistor of 
Fig.~\ref{fig-system}(c) as a function of the temperature of the gate. The different curves correspond to different
values of the gate thickness $\delta_3$. (b) The corresponding absolute value of the heat flux lost by the source.
(c) The corresponding net flux received or lost by the gate. (d) The corresponding amplification factor. All the 
results were obtained for a vacuum gap size of $d=500$ nm, and the source and drain temperatures are $T_\mathrm{S} = 20$ K 
and $T_\mathrm{D} = 1$ K, respectively. }
\label{fig-trans}
\end{figure}

In Fig.~\ref{fig-trans}(a-c) we present the different powers gained or lost by the three thermal reservoirs (drain, source 
and gate) as a function of the gate temperature $T_\mathrm{G}$ for different values of the gate thickness $\delta_3$ and a 
Nb-Au gap $d = 500$ nm. As expected, the heat received by the drain, $|Q_\mathrm{D}|$, increases monotonically with 
$T_\mathrm{G}$, irrespective of whether the gate is in the normal or in the superconducting state, see Fig.~\ref{fig-trans}(a). 
On the other hand, and contrary to what one could naively expect, the heat lost by the source, see panel (b), and the net heat 
in the gate, see panel (c), do not decrease monotonically upon increasing $T_\mathrm{G}$. This peculiar behavior only takes place
in the superconducting state ($T_\mathrm{G} < T_\mathrm{C}$) and it is a necessary condition for the amplification to occur, 
see Eq.~(\ref{eq-alpha}). In Fig.~\ref{fig-trans}(d) we show the corresponding value of the amplification factor, defined in 
Eq.~(\ref{eq-alpha}), as a function of $T_\mathrm{G}$. Notice that for thin Nb layers, values of $\alpha$ larger than 1 are 
possible in a certain range of temperatures within the superconducting phase, and for very specific values of $T_\mathrm{G}$ 
the amplification factor can reach values close to 100. In those cases, our three-body system truly behaves as a near-field
transistor. This is, however, never the case when the Nb gate is in its normal state where $\alpha$ tends to $1/2$.  

\begin{figure}[t]
\includegraphics[width=\columnwidth,clip]{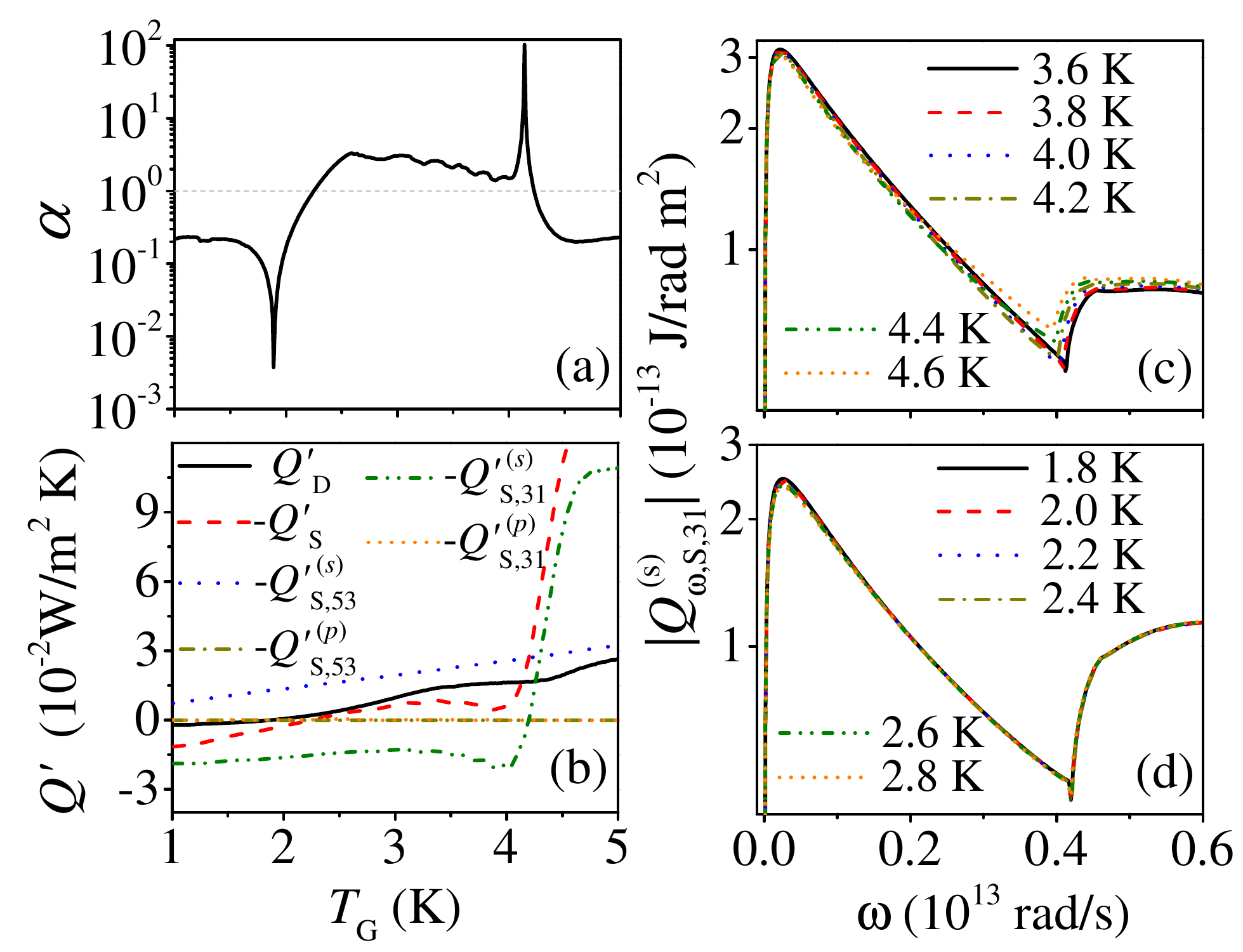}
\caption{(a) Amplification factor for a Nb gate with a thickness of 5 nm, separated from both the source and the drain by a 
500 nm vacuum gap, see Fig.~\ref{fig-trans}(d). (b) The corresponding derivates of the heat received by the drain, 
$Q^\prime_{\rm D}$, and of the heat lost by the source, $Q^\prime_{\rm S}$. Also included are the derivates of each 
one of the terms contributing to the heat lost by the source, see Eq.~(\ref{eq-Q_S}). (c) Absolute value of the spectral 
heat flux $Q_{\omega,{\rm S},31}^{(s)}=\Theta_{31}\tau_s ^{31}$ for several temperatures around the maximum 
amplification region in panel (a). (d) The same as in panel (c) for a gate of thickness 10 nm, in the temperature 
region where the corresponding amplification reaches its maximum in the Fig.~\ref{fig-trans}(d).}
\label{fig-der}
\end{figure}

To get some insight into the origin of the amplification, we show in Fig.~\ref{fig-der}(b) the derivatives of the heat 
exchanges in the source and drain with respect to the gate temperature, $Q^{\prime}_\mathrm{S/D}$, for the case of $\delta_3 
= 5$ nm shown in Fig.~\ref{fig-trans}(d), see also Fig.~\ref{fig-der}(a). From Eq.~(\ref{eq-alpha}) we see that amplification 
($\alpha > 1$) requires that $1 < Q^{\prime}_\mathrm{S}/Q^{\prime}_\mathrm{D} < 2$. The dependence with the gate temperature
shown in Fig.~\ref{fig-der}(b) explains why there is amplification in the range between approximately 2.5 and 4.2 K and 
why there is a huge peak at around 4.2 K, which corresponds to the situation in which $Q^{\prime}_\mathrm{S} \approx
Q^{\prime}_\mathrm{D}$. Moreover, we show in Fig.~\ref{fig-der}(b) the different individual contributions to $Q^{\prime}_\mathrm{S}$ 
following Eq.~(\ref{eq-Q_S}). There we see that the key temperature dependence of $Q^{\prime}_\mathrm{S}$ is mainly determined 
by the term involving $\Theta_{31} \tau_s^{31}$ for TE (or $s$) polarization, which is related to the heat lost by the source
to the combined gate-drain system. To understand the $T_\mathrm{G}$-dependence of this term, we illustrate in Fig.~\ref{fig-der}(c)
the evolution with the gate temperature of the corresponding spectral contribution. As one can see, the temperature dependence
is only significant in the frequency region close to $\omega_0 = 2\Delta_0/\hbar$ for the temperatures in which the amplification 
occurs. As we show in Fig.~\ref{fig-der}(d), that temperature dependence is much weaker when the gate thickness is increased, 
which explains why the amplification gets lost as the gate thickness increases. 

\begin{figure}[!t]
\includegraphics[width=\columnwidth,clip]{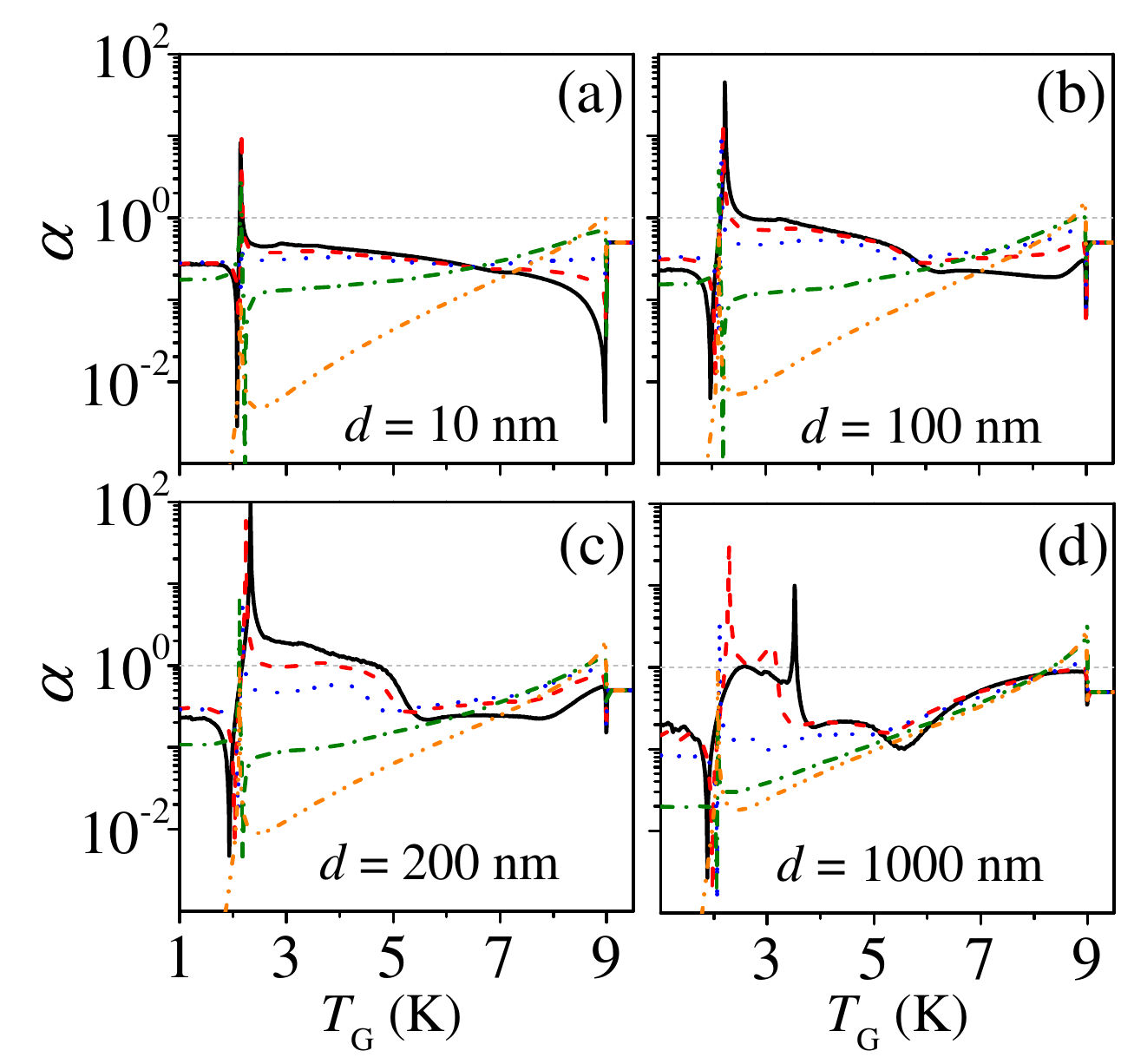}
\caption{Amplification factor in the thermal transistor of Fig.~\ref{fig-system}(c) as a function of gate temperature.
The different panels correspond to different vacuum gap sizes, while the different curves in each panel correspond to
different values of the gate thickness, $\delta_3$, whose values are the same as in Fig.~\ref{fig-trans}. In all cases 
the source and drain temperatures are $T_\mathrm{S} = 20$ K and $T_\mathrm{D} = 1$ K, respectively.}
\label{fig-amp}
\end{figure}

For completeness, we have systematically explored the role of the gap size $d$ between the Nb gate and the Au reservoirs.
We show a summary of these results in Fig.~\ref{fig-amp} where we show the amplification factor as a function of the gate
temperature for different values of $d$ (and also different values of the gate thickness). As one can see, the optimal 
amplification occurs for gap sizes $d$ on the order of a few hundred nm, while it only appears at some special points of 
the parameter space for smaller and larger gap values. In all cases, the amplification only appears for very thin Nb films. 

\section{Conclusions} \label{sec-conclusions}

Motivated by recent theoretical proposals and experiments on the use of phase-transition materials to realize near-field
thermal devices that mimic electronic components, we have presented in this work a theoretical study of the performance
of near-field thermal diodes and transistors based on the combination of superconductors and normal metals. We have shown 
that the drastic reduction in the thermal emission of a metal when it undergoes a superconducting transition can be utilized
to realize heat rectification in two-body systems and heat amplification in three-body devices. In particular, we have shown 
that a system formed by two parallel plates made of Au and (superconducting) Nb can act as a near-field thermal diode that 
exhibits striking rectification ratios very close to unity in a wide range of gap size values, outperforming all the existent 
proposals to date. Moreover, we have shown that a Nb layer placed in the middle of two Au plates can amplify the heat transferred 
to the drain at temperatures below Nb critical temperature. The ideas put forward in this work can be extended to propose other 
functional devices such as thermal logic gates and, overall, it illustrates the potential of the use of superconducting materials 
for near-field thermal management at low temperatures.

\acknowledgments

J.C.C.\ acknowledges funding from the Spanish Ministry of Economy and Competitiveness (MINECO) (contract No.\ FIS2017-84057-P).

\end{document}